\documentclass[11pt,twoside]{article}

\usepackage{asp2004}
\usepackage{epsf}
\usepackage{graphicx}
\usepackage{lscape}
\usepackage{bm}

\begin{document}
\title{The Characteristic Length Scale of the Magnetic Fluctuation in 
a Sunspot Penumbra: A Stochastic Polarized Radiative Transfer Approach}

\author{T. A. Carroll and J. Staude}
\affil{Astrophysikalisches Institut Potsdam, An der Sternwarte 16, D-14482 Potsdam, Germany}

\begin{abstract}
The characteristic size of penumbral structures are still below the
current resolution limit of modern solar telescopes. Though we
have seen a significant progress in theoretical work over the last
decades no tight constraints can be placed on the size of
penumbral structures in order to favor models with relatively
large and thick magnetic flux elements, just at or below the current
resolution limit, or on the other hand, clusters of optically
thin micro-structures.
Based on a macroscopic 2-component inversion and the approach of 
polarized radiative transfer in stochastic media, we have estimated 
the characteristic length scale of the magnetic fluctuation in a 
sunspot penumbra from observed Stokes spectra. The results yield a 
coherent picture for the entire magnetic neutral line of the penumbra
and indicate that the magnetic fluctuations
have a typical length scale between 30\,km and 70\,km.
\end{abstract}

\section{Introduction}
Solar sunspots are the most prominent manifestation of
concentrated magnetic fields on the solar surface and have been the
subject of an exhaustive amount of research. Despite the fact that the
photospheric structure and morphology is well studied and
described, and despite the advances in theoretical work \citep[see
e.g.][]{c2 TW04} there is no generally accepted picture available
which can explain the formation, structuring, and dynamics of
sunspots. It is generally accepted that one of the keys to increase our 
overall understanding of sunspots lies in a better
understanding of the many observable fine-scale features. 
The penumbra with its convectively driven small-scale
filamentary structure provides an excellent test-bed for any model
of magneto-convection. A wealth of observations and modelling
techniques gave us a good insight into the overall pattern 
and geometry of the magnetic field and plasma flows in the penumbra. 
The improvements in resolution and polarimetric sensitivity in the 
recent years has revealed an increasing number of
small-scale variations within the penumbral magnetic field (strength and
inclination), velocity, and temperature structures \citep[see
e.g.][]{c2 Sch02b}. Recent imaging observations have even
revealed fibrils of 150 to 180\,km in width which contain
substructures in the form of dark narrow cores, and which are less
than 90\,km in size \citep{c2 Sc02}. A small magnetic
structuring on sub-resolution scales (perpendicular and along the
LOS) has also long been suggested and recognized
by the ubiquitous Stokes profile asymmetry
\citep{c2 SM93,c2 Sch02a,c2 Mu02,c2 Tr04,c2 SA05}. Based on
the specific signatures of unresolved magnetic field structures
in the Stokes profiles, we exploit the diagnostic capabilities of 
the meso-structured approach proposed by
\citet{c2 CS03,c2 CS05,c2 CS06}, which is based on a stochastic
formulation of the polarized radiative transfer. This new approach
is applied to spectro-polarimetric observations of a sunspot
penumbra which were subject to a macroscopic 2-component
inversion by \citet[][hereafter RBC]{c2 BR04}. Using this
combination of macroscopic and mesoscopic inversion
allows us to estimate the characteristic length scale of the 
underlying magnetic fluctuation.

This paper is organized as follows: In Sect.\ 2 we give a brief
overview of the basic concepts of the line formation in
fluctuating and stochastic media with structures of finite length.
In Sect.\ 3 we demonstrate that the statistical scattering and
absorption terms of the stochastic transfer equation directly
determine the degree of the area and amplitude asymmetries of the
resulting Stokes-$V$ profiles, which can be used to estimate the
typical length scale of the magnetic fluctuation. In Sect.\ 4 we
describe the basic results of the 2-component inversion and the
following application of the meso-structured approach. Finally, in 
Sect.\ 5 we summarize our results.

\section{The Stochastic Transfer Equation for Polarized Light}
\label{c2 sec:stochtrans} 
Any approach to radiative transfer modelling has to make assumptions 
about the underlying atmospheric structure. In atmospheres which
are composed of small scale fibril structures -- smaller than the resolution
element -- the natural question arises how these structures can be 
adequately accounted for in polarized radiative transfer modelling?
The two (magnetic) modelling approaches
based on a macro-structured (conventional flux-tube model) or
micro-structured \citep[MISMA,][]{c2 SA96} description of the
atmosphere cannot deal with a-priori unknown structures 
of finite spatial extent.
To describe the polarized radiative transfer in an arbitrary fluctuating
atmosphere, we pursue a rigorous probabilistic formulation of the
underlying atmospheric structure.
A detailed description of that stochastic approach
and the derivation of the stochastic transport 
equation for polarized light, is given in \citet{c2 CS05,c2 CS06}. 
In this contribution we will only summarize the basic concepts of 
this approach which puts forward the idea of a more general 
meso-structured magnetic atmosphere
\citep[MESMA,][]{c2 CS06}.

A prerequisite for a stochastic modelling of structures with
finite spatial extent is to take into account the complete hierarchy of
spatial correlations.
If we assume that the spatial coherency or correlation
of the individual atmospheric components along a given trajectory
can be approximated by a Markov process, all the
higher spatial correlations can be reduced to first-order correlation effects.
We therefore introduce a random atmospheric
vector $\bm{B}$ which comprises all relevant atmospheric parameters 
such as temperature,
pressure, velocity, magnetic field strength, magnetic field
inclination, etc., and specify the spatial
correlation of the individual atmospheric components by the following
discontinuous Markov process (the space-dependent, conditional-probability 
density function) 
\begin{equation}
p(\bm{B}'',s+\Delta s\,|\,\bm{B}',s) = e^{-\gamma(\bm{B}') \Delta s}
\; \delta(\bm{B}' -\bm{B}'') +
\left[1 - e^{- \gamma(\bm{B}') \Delta s}\right] p(\bm{B}'')\;,
\label{c2 kubo}
\end{equation}
where $\gamma(\bm{B}')$ is the fluctuation rate
of the atmospheric structure 
$\bm{B}_s$ and $p(\bm{B}'')$ the spatial independent (stationary) 
probability density.  The fluctuation rate can be expressed with the help 
of the correlation length $l$ (characteristic length scale) to give
\begin{equation}
\gamma(\bm{B}_s) = l^{-1}(\bm{B}_s)\;.
\label{c2 corrlength}
\end{equation}
This discontinuous Markov process, also known under the name
Kubo-Anderson process, was already used by \citet{c2 FF76} to
describe inhomogeneous velocity fields in stellar and solar
atmospheres. This process describes the spatial relationship in
terms of the conditional probability density. 
$p(\bm{B}'',s+\Delta s\,|\,\bm{B}',s)$. It gives the probability of 
being in the state
or atmospheric regime $\bm{B}''$ at the spatial position $s+\Delta
s$, after having moved a short distance $\Delta s$ from the known
atmospheric state $\bm{B}'$ at $s$. The probability for staying in
the initial regime $\bm{B}'$, expressed by the Dirac delta
function, decays exponentially with $\Delta s$ while a sudden jump
into another regime $\bm{B}_{s+\Delta s}$ has an exponentially
increasing probability and is only weighted by the overall stationary
probability density of the final state $\bm{B}''$. This
conditional probability density function thus describes the
correlation between the two spatial positions $s$ and $s+\Delta s$
along a given trajectory (later the LOS). The degree of
correlation is controlled by the fluctuation rate, $\gamma$, or 
the correlation length, $l$, respectively, which describe the mean
path length or extent of the atmospheric structure. This generic
stochastic model of the atmospheric structure allows us to
formulate a differential equation for the depth-dependent
atmospheric probability density function (PDF) $p(\bm{B},s)$ 
from which we can derive a stochastic transport equation 
for polarized light
\citep[see][for a detailed derivation]{c2 CS05,c2 CS06}. 
For the particular application in this contribution, an appropriate
model may be given by the discontinuous Kubo-Anderson process 
Eq.~(\ref{c2 kubo}), which allows to derive a master-like transport
equation for the mean conditional Stokes vector, which reads
\begin{equation}
\frac{\partial \bm{Y_{B}}}{\partial s} =  -\bm{K_B}\bm{Y_{B}} + \bm{j_B} +
\int \bar{\gamma}(\bm{B}) \bm{Y_{B'}}\,p(\bm{B}',s)\;d\bm{B}' 
- \int \gamma(\bm{B}')
\bm{Y_{B}}\,p(\bm{B}')\;d\bm{B}'\;,
\label{c2 meancond}
\end{equation}
where $\bar{\gamma}$ is the modified fluctuation rate
\begin{equation}
\bar{\gamma}(\bm{B}) = \frac{p(\bm{B})}{p(\bm{B},s) l(\bm{B})} \: ,
\label{c2 modgamma}
\end{equation}
and $\bm{Y_B}$ is the mean conditional Stokes vector which is defined as
\begin{equation}
\bm{Y_B}(s) = \int \bm{I}\,p(\bm{I},s\,|\,\bm{B},s) \; d\bm{I}\;.
\label{c2 meanconddef}
\end{equation}
Please note, that $p(\bm{B})$ and $p(\bm{B},s)$ are in general not the same 
distributions, the latter is the spatial dependent pdf of the
atmospheric state vector $\bm{B}$ which evolves
from a given initial state along the ray path, whereas the 
stationary pdf, $p(\bm{B})$, is an a priori assigned attribute of the atmosphere.
The evolution of $p(\bm{B},s)$ is governed by a Master equation 
\citet{c2 CS05,c2 CS06}. 
If the atmosphere is structured on finite scales, $p(\bm{B},s)$
eventually converges to the stationary distribution $p(\bm{B})$,
the equilibrium state.

The stochastic transport equation for the mean conditional Stokes
vector has a rather simple physical interpretation. Since
$\bm{Y_{B}}$ is conditioned on a particular atmospheric regime
$\bm{B}$, and the transport equation (\ref{c2 meancond}) describes its
statistical evolution through the atmosphere, four physical
processes govern the transport of the mean conditional Stokes
vector: the two processes of true absorption and thermal emission,
and two processes that describe the statistical inflow and outflow
of intensity to or from the regime $\bm{B}$ under consideration.
In detail, the third term on the r.h.s.\ of Eq.~(\ref{c2 meancond})
gives the amount of intensity or photons which may enter from all
other atmospheric regimes into $\bm{B}$ and can therefore be
considered as an additional (statistical) source or scattering
vector, while the fourth term describes the (statistical) loss or
absorption of intensity due to transitions of photons from
$\bm{B}$ to all other possible atmospherical regimes. The degree
of statistical scattering and absorption is controlled by the
correlation length $l$ of the particular atmospheric regime. The
observable of our problem---the expectation value of the Stokes
vector at the top of the atmosphere---can easily be obtained from
a final integration of the mean conditional Stokes vector over the
entire atmospheric state space of $\bm{B}$,
\begin{equation}
\langle\bm{I}(s)\rangle = \int \bm{Y_B}(s)\,p(\bm{B},s)\;d\bm{B}\;.
\label{c2 meanstokes}
\end{equation}

\begin{figure}[!t]
\centering
\includegraphics[height=46mm]{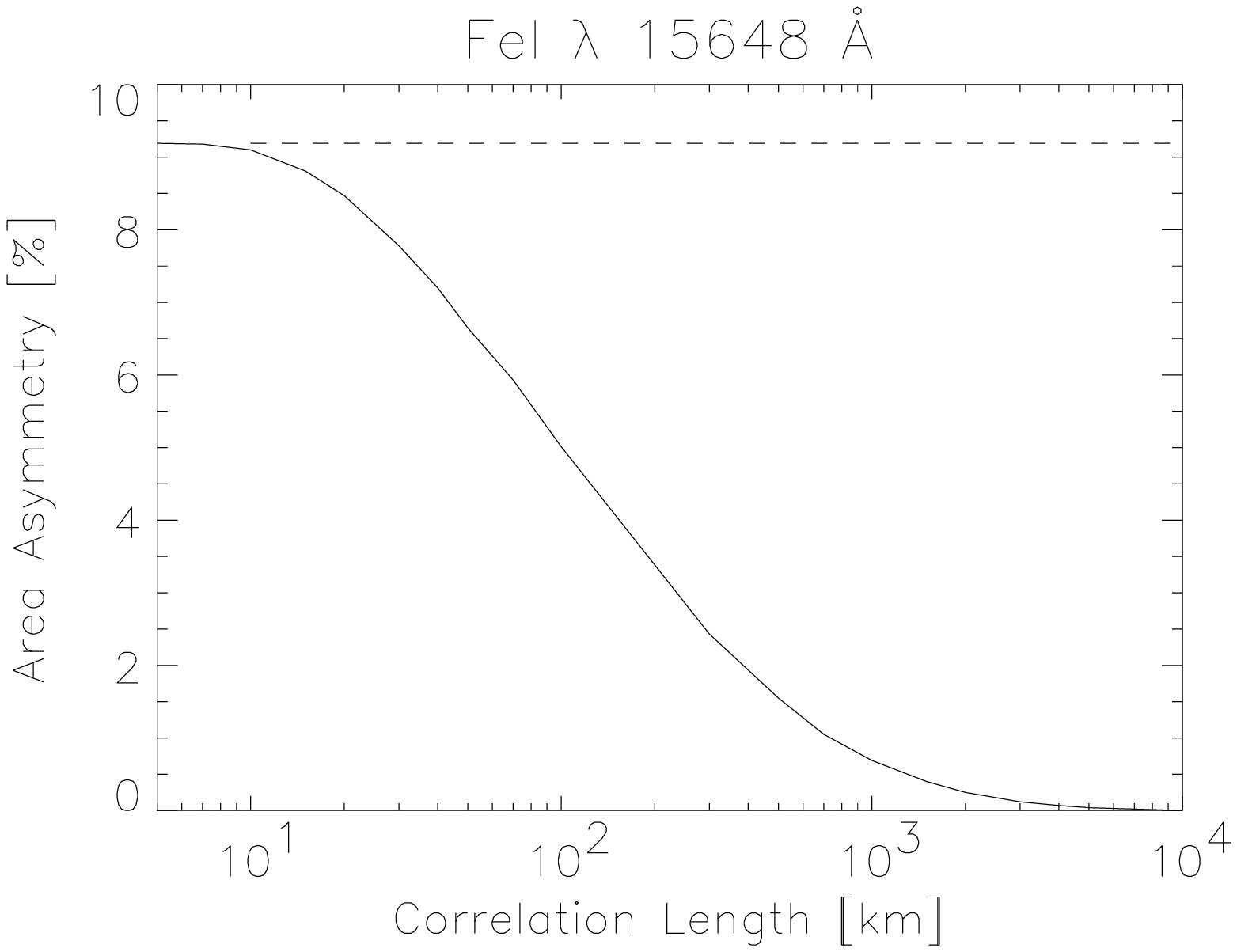}\hspace{1pt}
\includegraphics[height=46mm]{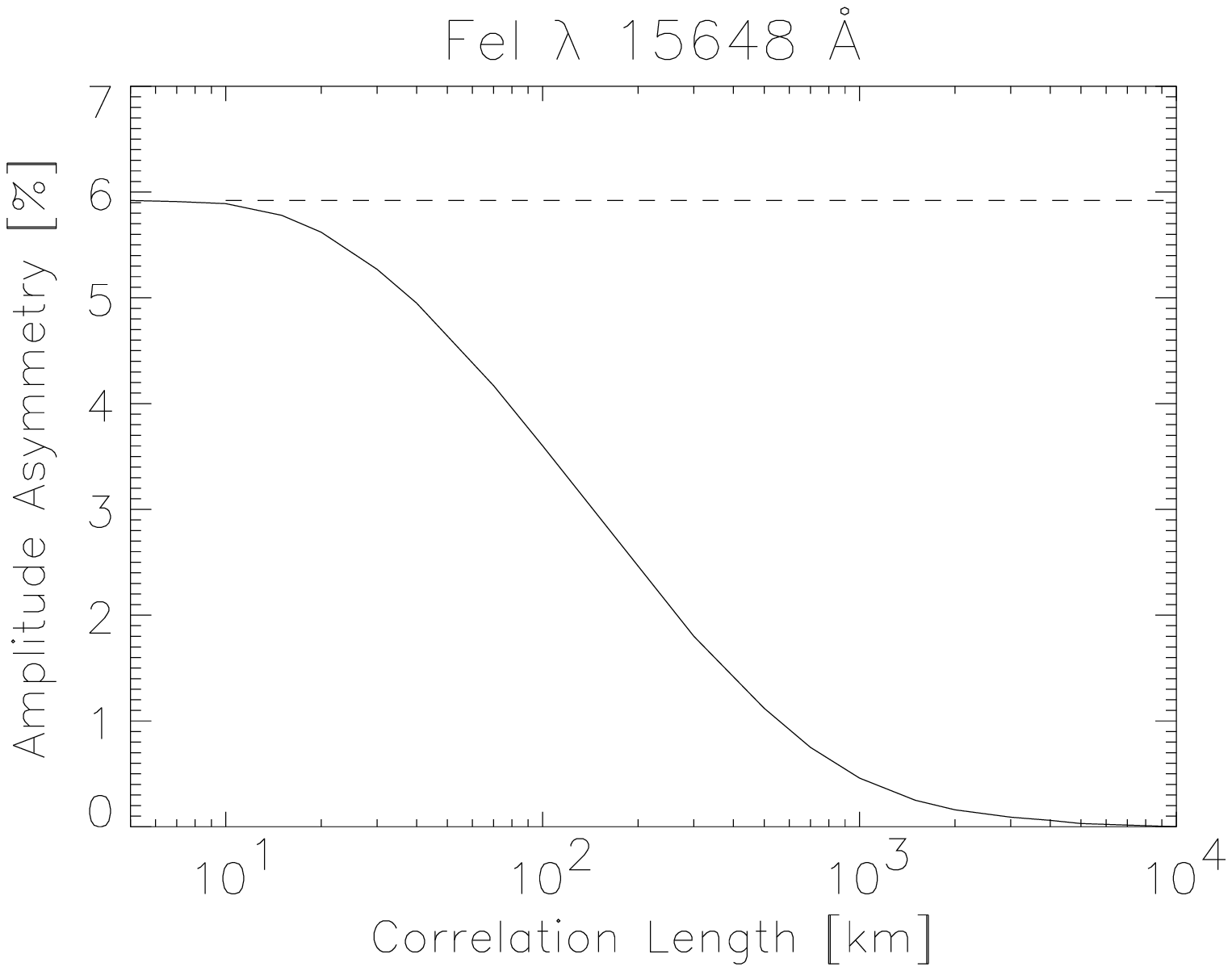}
\caption{Left: Area asymmetry vs correlation length.
Right: Amplitude asymmetry vs correlation length. The 
area and amplitude asymmetries reveal a clear functional dependence 
on the correlation length. The dashed lines in both figures indicate the
area and amplitude asymmetries which would originate from a pure
micro-structured atmosphere, calculated under the MISMA
approximation.} \label{c2 fig:area_ampl}
\end{figure}

\section{Stokes Profile Asymmetries}
\label{c2 sec:asymmetries} Stokes-$V$ profile asymmetries can be
attributed to the existence of gradients in velocity and magnetic
fields \citep{c2 LL96,c2 LA02}. Thus the
net circular polarization (NCP) and area asymmetry are good
indicators for the underlying magnetic inhomogeneity along the
LOS. \Citet{c2 CS06} show that in a magnetic fluctuating or randomly
organized atmosphere the statistical scattering term in 
Eq.~(\ref{c2 meancond}) is responsible for most of the NCP. To
demonstrate this effect, we performed some simple model calculations 
for the \ion{Fe}{i} line at 1564.8\,nm where we
have assumed a stochastic 2-component atmosphere, which consists
of a field-free ensemble of structures with a net downflow of 
1.5\,km/s, and a stationary magnetic ensemble of structures with a field
strength of 700\,G. To quantify the degree of asymmetry we 
used the usual definitions of the area asymmetry, $\delta A$, and
the amplitude asymmetry, $\delta a$,
\begin{equation}
\delta A=s\,\frac{\int_{\hat{\lambda}} V(\lambda)\,
d\lambda}{\int_{\hat{\lambda}}\,|\, V(\lambda) \,|\,d\lambda}\;,
\qquad
\delta a = s\,\frac{\max\{V(\lambda)\} +
\min\{V(\lambda)\}}{\max\{V(\lambda)\} - \min\{V(\lambda)\}}\;,
\end{equation}
where s is the sign of the blue lobe of the Stokes-$V$ profile.

\begin{figure}[t]
\centering
\includegraphics[height=44mm]{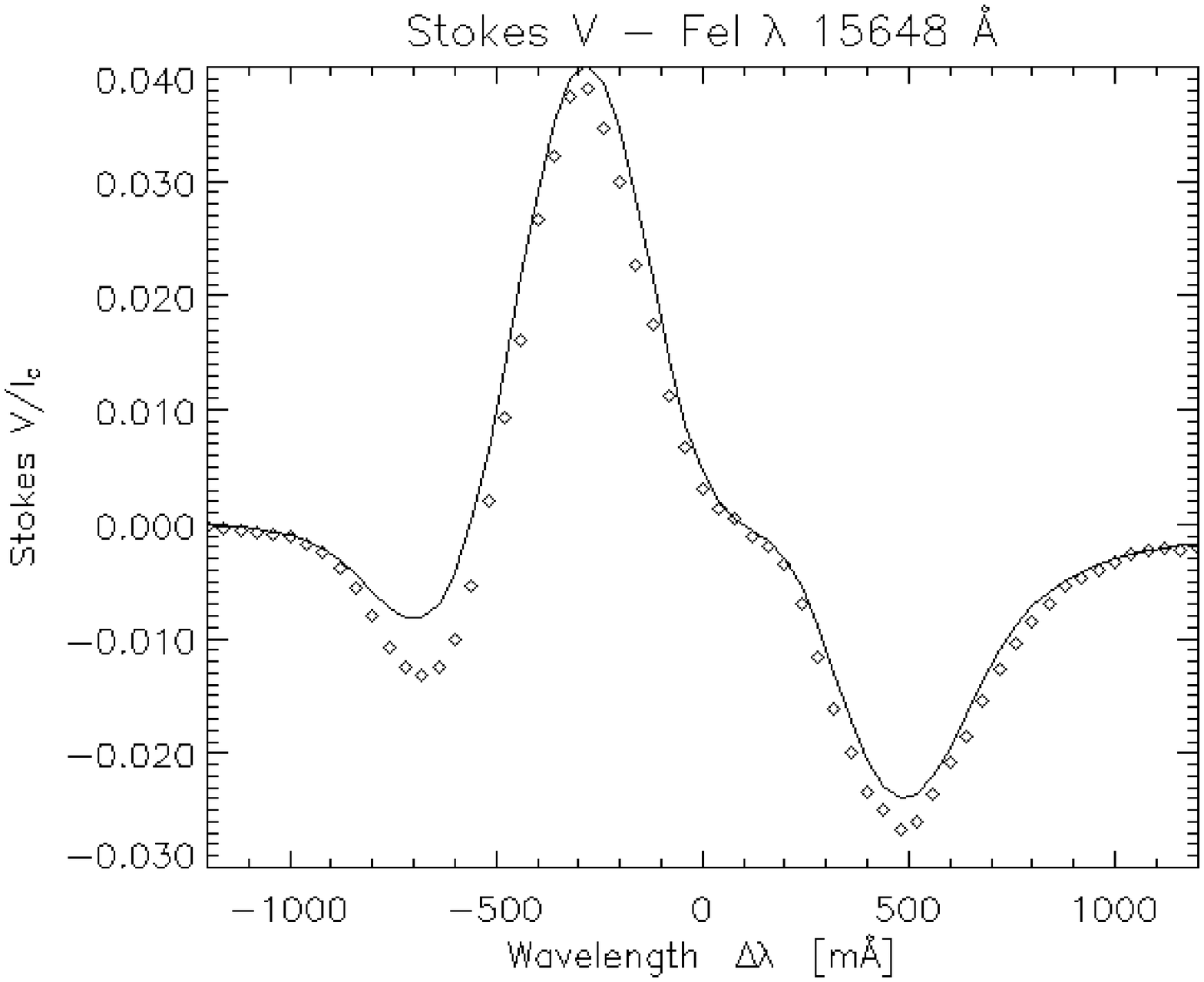}\hspace{5pt}
\includegraphics[height=44mm]{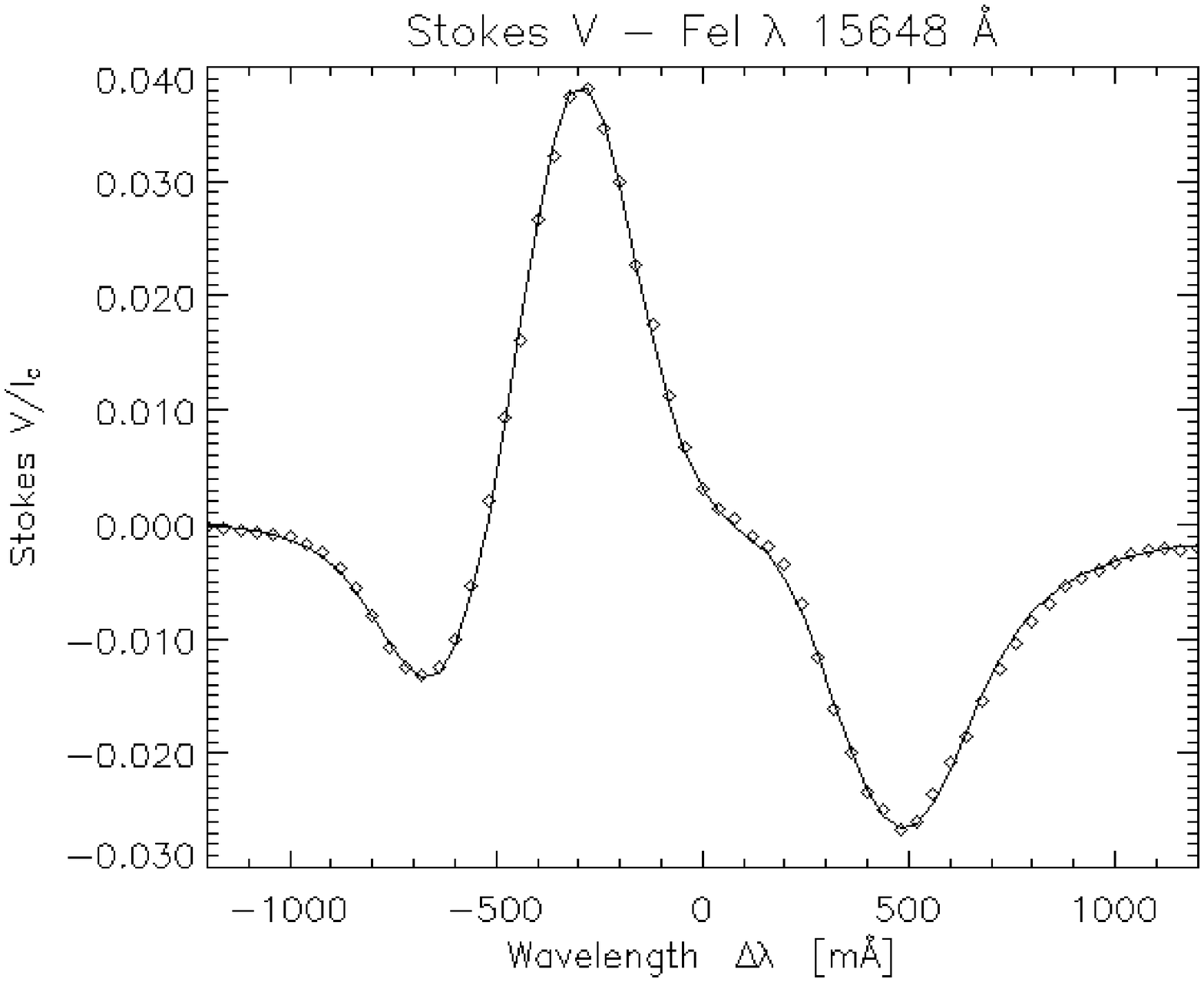}
\caption{Left: The averaged Stokes-$V$ profile from the
2-component inversion (in the macroscopic limit) by \citet{c2 BR04}.
Right: The averaged Stokes-$V$ profile after adjusting the correlation
length to 40\,km (diamonds: observations; solid lines: fitted
profiles).} \label{c2 fig:init_best}
\end{figure}

Figure \ref{c2 fig:area_ampl} shows the Stokes-$V$ profile area and
amplitude asymmetry for \ion{Fe}{i} 1564.8\,nm as a function of the 
correlation length. Figure \ref{c2 fig:area_ampl}
nicely demonstrates, how the area and amplitude asymmetries
depend on the correlation length (characteristic length scale) of
the underlying atmospheric structures. 
The smaller the correlation lengths, or higher the
fluctuation rates, the stronger the asymmetries.
Please note, that only the correlation length has changed in that
scenario. For very small correlation lengths the asymmetries
asymptotically reach the micro-turbulent limit ($l \rightarrow
0$). For a better comparison of the asymptotical behavior the \emph{true} 
micro-turbulent limit was calculated under the MISMA approximation and is
denoted by the dashed line in Fig.~\ref{c2 fig:area_ampl}. For very
large correlation lengths and low fluctuation rates (the macrostructured limit)
the asymmetries asymptotically approach zero. 
This is the expected result as the magnetic and non-magnetic
structures in our model calculations possess no intrinsic gradients. 
This behavior is also a direct consequence of the stochastic transport
equation (\ref{c2 meancond}) where the statistical scattering and
absorption are inversely proportional to the correlation length
$l$, and in the limit $l \rightarrow \infty$ the different
atmospheric regimes finally decouple and no statistical scattering
and absorption can occur. \Citet{c2 CS06} show that the stochastic 
transfer equation presented here includes, in fact, the micro-structured
(MISMA) approximation as well as the macro-structured approach as
limiting cases. Please note, that already for correlation
lengths greater than 10\,km significant deviations from the
micro-turbulent approximation occur. As the inherent magnetic
fluctuation of the atmosphere has an important effect on the
degree of the NCP or area asymmetries \citep{c2 CS06}, these
asymmetries can be used in turn to estimate the underlying length
scale of the magnetic fluctuation.

\section{Penumbral Structures from a Stochastic 2-Component Approach}
\label{c2 sec:analysis} The starting point of our analysis are the
results of a 2-component inversion of a sunspot penumbra from
RBC. Their retrieved overall penumbral
model is in agreement with a discontinuous model of the so-called
uncombed penumbra proposed by \citet{c2 SM93}, with an almost
horizontal flux-tube component embedded in a more vertical
background field component. The inversion were performed by a
special adaptation of the SIR code (see RBC for details) and
returns the temperature stratification, the three components of
the single-valued magnetic field vector, and the LOS velocity of the two
components of the model, together with a single value for the
macro-turbulent velocity. In addition, the stray light factor
$\alpha$ and the filling factor of the flux-tube component are
also determined. We do not go into the details of the analysis
and results of the inversion of RBC, but we want to emphasize that
the quality of the fitted profiles of the 2-component inversion seems 
to be able to grasp the essential character of the underlying 
penumbral atmosphere, despite the fact that the magnetic field and 
velocity structures of the two components are assumed to be neither
depth dependent nor interlaced. The 2-component inversion of RBC 
represents therefore a typical macro-structured approach, because
the two components are not interlaced along the LOS 
and, unlike the observed Stokes-$V$ profiles, the resulting
synthetic profiles therefore exhibit no NCP or area asymmetry. In 
order to allow for a more flexible meso-structured approach we have 
incorporated the model parameters of the 2-component atmospheres into 
our stochastic (meso-structured) transfer model.
The underlying atmosphere in this stochastic scenario can now be
considered as the composition of two different ensemble structures 
(the background and the flux-tube ensemble) but this time they are
really interlaced.
The mean length scale or extent of the individual
ensemble structures is determined by the correlation length, which at the same
time determines the rate of the fluctuation within the atmosphere
and the \emph{ability} of the atmosphere to produce a NCP.
For each individual average profile the
correlation length of the underlying structures were adjusted 
until a best fit with the observation were reached. 
The profiles were averaged over
a small portion of pixels ($3\times3$). We put particular
emphasis on profiles along the magnetic neutral line because of
the preferred viewing angle onto the magnetic structures in this region.
Figure \ref{c2 fig:init_best} shows an averaged Stokes-$V$ profile located
in the magnetic neutral line. Despite the fact that the initial model
parameters of the 2-component inversion are relatively good
(Fig.~\ref{c2 fig:init_best}, left), they cannot account for the
area asymmetry of the observed profile. The observed Stokes-$V$ profile
of \ion{Fe}{i} 1564.8\,nm has a NCP of 4.9\,m\AA\ and an area 
asymmetry of $-17.6$\%. Since the infrared iron line is not very 
sensitive to velocity and magnetic-field gradients or fluctuations, 
due to its weakness and small range of formation, the observed profiles
must be subject to a rather large fluctuation rate. After 
adjusting the correlation length (with all other parameters kept
fixed), the fit of the profiles could be considerably improved and
the resulting NCP of 4.9\,m\AA\ and area asymmetry of $-17.3$\%
could be almost exactly reproduced. To obtain these large values
of the Stokes-$V$ asymmetries, with the given model parameters of
the 2-component inversion, a correlation length as low as 40\,km
was needed. The analysis of a number of Stokes-$V$ profiles along
the magnetic neutral line gave similar results. For each of the analyzed
Stokes-$V$ profiles we could obtain a significantly improved fit
compared to the macroscopic analysis of RBC, simply by varying the
underlying correlation length. From our analysis we could infer that
the typical length scale of the magnetic fluctuation in a
penumbra must be between 30 and 70\,km.

\section{Summary}
\label{c2 sec:summary}
The stochastic polarized radiative transfer 
recently developed by \citet{c2 CS05,c2 CS06} was applied to a
meso-structured magnetic model \citep[MESMA,][]{c2 CS06} of a
sunspot penumbra to estimate the characteristic length scale of
the magnetic fluctuation. The degree of fluctuation directly
controls the amount of the NCP (or area asymmetry) of the
resulting circular polarization signal. A maximum degree of
asymmetry is always produced by a micro-structured or
micro-turbulent atmosphere, whereas the NCP vanishes for a
macro-structured atmosphere. However, as it can be seen from
Fig.\ \ref{c2 fig:area_ampl}, these two cases only represent the very far
limits of a more general meso-structuring that is valid for the broad
range of finite length scales. Based on the results of 2-component 
analysis of a sunspot penumbra (RBC),
we applied a method which
is able to account for the finite fluctuation of magnetic
components along the LOS. With this approach, we were able to
estimate the correlation length of the underlying magnetic
structures, for a number of locations along the magnetic neutral
line. For the characteristic length scales of the magnetic fluctuation 
we found values between 30 and 70\,km. These findings
indicate that the magnetic field in the penumbra, although
it may have on average a typical uncombed-like structure,
may possess a considerable amount of substructures. We conclude that
the meso-structured analysis presented here,
may provide a viable approach for studying small-scale and complex
magnetic fields in the solar atmosphere.

\acknowledgments
The authors are grateful to L.\ R.\ Bellot Rubio, H.\ Balthasar, and 
M.\ Collados for providing the data of their 2-component inversion. 
We also gratefully acknowledge financial support by the Deutsche
Forschungsgemeinschaft (DFG) under the grant CA 475/1-1.

\end{document}